\documentstyle[twocolumn]{jpsj}

\def\runtitle{
Antiferromagnetic Quantum Spins on the Pyrochlore Lattice}
\def\runauthor{Hirokazu {\sc Tsunetsugu}}

\title
{
Antiferromagnetic Quantum Spins on the Pyrochlore Lattice
}

\author
{ 
Hirokazu {\sc Tsunetsugu}
}

\inst
{
Institute of Materials Science, University of Tsukuba, 
Tsukuba 305-8573   
}

\recdate
{
Dec.~25, 2000
}

\abst
{
The ground state of the $S={1 \over 2}$ Heisenberg antiferromagnet 
on the pyrochlore lattice is theoretically investigated.  
Starting from the limit of isolated tetrahedra, I include 
interactions between the tetrahedra and obtain an 
effective model for the spin-singlet ground state multiplet 
by third-order perturbation.  
I determine its ground state using the mean-field approximation 
and found a dimerized state with a four-sublattice structure, which 
agrees with the proposal by Harris et al.  
I also discuss chirality correlations and spin correlations 
for this state.  
}

\kword
{
geometrical frustration, Heisenberg antiferromagnet, pyrochlore lattice, 
dimerization, chirality
}

\begin{document}
\sloppy
\maketitle

\input psbox.tex 

Antiferromagnetic spin systems on geometrically frustrated 
lattices show many unusual behaviors in magnetic and thermal 
properties.~\cite{Liebmann,Ramirez}  
Their origin may essentially be attributed to 
the point that their 
classical ground state has a thermodynamic 
degeneracy with a potential instability 
towards various phases.  Villain pointed out the possibility that thermal 
fluctuations drive symmetry breaking to the 
phase with the largest entropy.~\cite{Villain}  
This mechanism of order by 
disorder is also expected for quantum fluctuations.  
In this paper, I study the ground state of 
the $S={1 \over 2}$ Heisenberg antiferromagnet 
on the pyrochlore lattice and examine the possibility of 
a new quantum phase.  

The pyrochlore lattice is a three-dimensional (3D) network 
of corner-sharing tetrahedra (see Fig.~1), realized as 
a sublattice of many compounds including pyrochlores 
and spinels.  
Since all the elementary closed paths of spin exchange processes 
are triangular, antiferromagnets on the pyrochlore lattice 
are subject to strong frustration due to this lattice 
geometry.  Another important character is the loose 
packing of the lattice, which is in contrast with the closed 
packing of the fcc lattice, another example of a geometrically frustrated 
lattice in 3D.  

The combination of geometrical frustration 
and loose packing of the pyrochlore lattice leads 
to thermodynamically degenerate ground states of 
classical antiferromagnets on it, which was first pointed out by 
Anderson for the Ising model.~\cite{Anderson}  Numerical simulations 
for the classical antiferromagnets indicate the absence 
of a finite-temperature magnetic phase transition,~\cite{Liebmann} 
suggesting that thermal fluctuations cannot break symmetry.  

Quantum fluctuations are 
another channel of lifting the classical ground-state 
degeneracy.~\cite{Nikuni}  
This possibility for the pyrochlore lattice 
was first examined by Harris {\it et al}.~\cite{Harris}  
They used a field-theoretical 
approach and degenerate perturbation for the spin-singlet 
ground-state multiplet and showed a sublattice ordering of dimerization.  
This was followed by Canals and Lacroix, who 
also started from decoupled tetrahedra and calculated 
the spin-spin correlation 
function at finite temperatures, using perturbation in 
the tetrahedron-tetrahedron interactions up to the second order.~\cite{Canals}  
The result shows that the spin correlation length remains very short 
at the lowest temperature, and they concluded that 
the ground state is a quantum spin liquid with a finite 
spin gap, which 
is consistent with the claim of Harris {\it et al}.  
In the pyrochlore lattice, it is believed that there exist a thermodynamic 
number of singlet states below the spin gap.  
In this paper, I study the possibility of symmetry breaking 
in the singlet sector driven by quantum fluctuations.  
To this end, I follow a similar approach as those of the previous two works 
and start from a set of weakly interacting tetrahedra.  
I construct a full effective model for the spin-singlet 
ground-state multiplet within the third-order degenerate 
perturbation.  I show that its mean-field solution 
is a partially dimerized state with a four-sublattice structure, 
and discuss spin and chirality correlations.

The pyrochlore lattice consists of two types of tetrahedra 
with the same number, say type A and B, and the network 
of these tetrahedra is bipartite as shown in Fig.~1.  
I study the Hamiltonian 
of the $S={1 \over 2}$ Heisenberg antiferromagnet on this lattice, 
\begin{figure}[b]
$$\psboxto(0cm;5cm){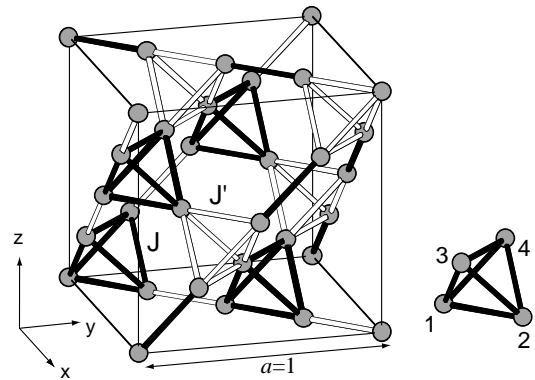}$$
\caption{Pyrochlore lattice and a unit of type-A tetrahedron.  
Tetrahedra of type A (B) are 
shown by black (white) bonds.
The lattice constant of the cubic unit cell is set 
to units of length.}  
\label{fig1}
\end{figure}
\begin{eqnarray}
H = J \!\!\!  \sum_{\langle {\bf r},{\bf r}' \rangle \in {\rm A}}  
 \!\!\! {\bf S}_{\bf r} \cdot {\bf S}_{{\bf r}'} + 
J' \! \! \! \sum_{\langle {\bf r},{\bf r}' \rangle \in {\rm B}}  
 \!\!\! {\bf S}_{\bf r} \cdot {\bf S}_{{\bf r}'}, 
\ (J,J'>0)
\label{eq:ham}
\end{eqnarray}
where the first and second summations are taken over bonds on 
type-A and B tetrahedra, respectively.  
To study the case of homogeneous coupling $J'$=$J$, 
I start from the limit of $J'$=0 and include $J'$-terms 
as perturbation.  

When $J'=0$, all type-A tetrahedra are decoupled 
from each other, and we can easily construct eigenfunctions 
of the entire system from those of one tetrahedron.  
On an isolated tetrahedron, 
eigenfunctions of its total spin $S_{\rm A}$ are also 
eigenfunctions of the Hamiltonian.  Since there are four spins each with 
$S={1 \over 2}$, $S_{\rm A}$ is either 0, 1, or 2, with 
degeneracy 2, 3, or 1, and energy $-{3 \over 2}J$, 
$-{1 \over 2}J$, or $+{1 \over 2}J$, respectively.  
In the ground-state multiplet, each type-A tetrahedron takes 
one of the two $S_{\rm A}$=0 states.  Therefore, 
the degeneracy is $2^{N_{\rm A}}$ ($N_{\rm A}$: the 
number of type-A tetrahedra, {\it i.e.}, a quarter of 
the number of sites), and all the states 
are of course spin singlet.  
The lowest excited multiplet is a spin triplet, and 
the spin gap is $J$.  

As we will see later, it is convenient to use a chiral basis 
for each tetrahedron.  
In the $S_{\rm A}^z=0$ subspace, two spin-singlet states are 
\begin{equation}
  \begin{array}[t]{l}
  |0,+\rangle = [ 
              (12)_+
   + e^{i\phi}   (13)_+
   + e^{-i\phi} (14)_+ ]/\sqrt{3}, \\ 
   |0,-\rangle = [ 
               (12)_+
    + e^{-i\phi} (13)_+
    + e^{i\phi}   (14)_+ ]/\sqrt{3}, 
  \end{array}
  \label{eq:base0}
\end{equation}
where 
$(ij)_{\pm}=
(S_i^- S_j^-|\!\!\uparrow \uparrow \uparrow \uparrow \rangle
\pm S_i^+ S_j^+ |\!\!\downarrow \downarrow \downarrow \downarrow \rangle 
)/\sqrt{2}$ and 
$\phi$=$2\pi/3$.  
As for the spin triplet, we obtain 
two chiral states, $|1,+\rangle,  |1,-\rangle$, by replacing 
$(ij)_+$ with $(ij)_-$, 
completed by the non-chiral state 
$|1,0\rangle = [ (12)_- + (13)_- + (14)_- ]/\sqrt{3}$.  The spin-quintet 
state is also non-chiral, 
$|2,0\rangle = [ (12)_+ + (13)_+ + (14)_+ ]/\sqrt{3}$.  
Let us define the spin chirality by 
$\chi_{ijk}={4 \over \sqrt{3}} 
{\bf S}_i \cdot ({\bf S}_j \times {\bf S}_k$), 
and these bases are also the eigenfunctions of the chirality around site $1$, 
$\chi_{234} |S,\tau \rangle = \tau |S,\tau \rangle$ ($\tau = +,-,0$).  
We can also define the chirality around other sites in a similar way, 
and it is important that the spin singlets do not change their 
chirality, while the triplets do.  
This is easily understood since the singlets 
belong to the complex 1D representations,  
$\Gamma_2$ and $\Gamma_3$, 
of the tetrahedral point group, 
whereas the triplets belong to the 3D representation T$_2$.  

\begin{figure}[bt]
$$\psboxto(7.5cm;0cm){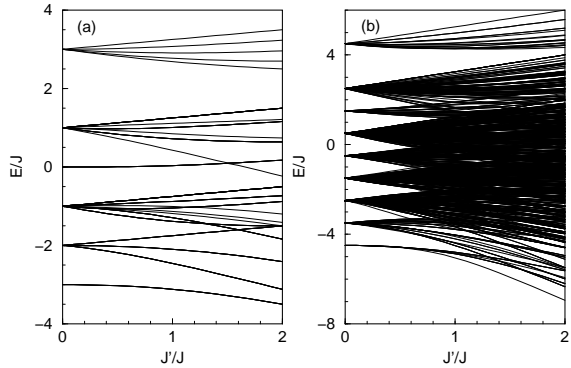}$$
\caption{Eigenenergies of the Heisenberg antiferromagnet on 
(a) a tetrahedron pair and (b) a tetrahedron triple.}
\label{fig2}
\end{figure}

When the $J'$-term is switched on, this leads to a mixing of 
different states in the ground-state multiplet, and we may 
expect a lift of the thermodynamic degeneracy.  
First, consider a pair of neighboring 
type-A tetrahedra, and connect them by a single bond with 
exchange coupling $J'$.  
I show in Fig.~2(a) 
the energy of 70 states in $S_{\rm tot}^z=0$ subspace, as a function 
of $J'$.  The degeneracy of the ground-state multiplet 
is 4-fold, and it is never lifted.  Therefore, we have to go 
to the next step, and consider three neighboring type-A tetrahedra 
each pair of which is connected by one bond.  There are two 
different cases.  In the first case, two perturbative bonds 
are connected to the same site on each tetrahedron, 
whereas in the second case they are connected to 
two distinct sites.  It is found that 
the second case lifts the degeneracy, but the first case does not.  
Figure 2(b) shows the energy of 924 states in the $S_{\rm tot}^z=0$ subspace, 
for the second case.  The ground-state multiplet with 8-fold degeneracy 
is now split into 4 levels with 1, 3, 1, and 3 fold degeneracies, 
respectively, in the ascending order of energy.  
At $J'=J$, this energy split is 0.2468$J$ 
between the lowest and highest levels, while 
the singlet-triplet gap is 0.5491$J$.  

\begin{figure}[tb]
$$\psboxto(0cm;4cm){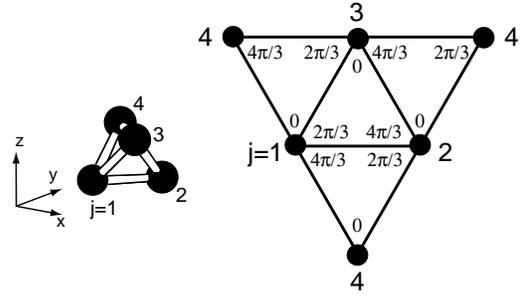}$$
\caption{Supertetrahedron comprised of four type-A 
tetrahedra, and value of parameter $\alpha_j$ 
in the effective model.}
\label{fig3}
\end{figure}

These behaviors are described by degenerate perturbation in $J'$.  
The second order only yields a constant self-energy, $-{9 \over 64} J'^2/J$ 
per spin, and we require third-order terms to mix different chirality 
configurations on a tetrahedron triple.  To describe chirality for 
the tetrahedron positioned at ${\bf R}$,  
I introduce spin-${1 \over 2}$ pseudospin operators, 
${1 \over 2}{\mib \tau}_{\bf R}$, 
such that their $z$-component is a diagonal matrix in the chirality space.  
Using these operators, the third-order effective model is written 
as~\cite{NB1} 
\begin{equation}
  H_{\rm eff}^{(3)} = 
  \begin{array}[t]{l}
    \displaystyle
  \sum_{\langle {\bf R}, {\bf R}', {\bf R}'' \rangle} 
  [ c - b 
  (1 - 2 {\bf e}(\alpha _1) \cdot {\mib \tau}_{\bf R}) 
  \\
    \displaystyle 
  \times 
  (1 - 2 {\bf e}(\alpha _2) \cdot {\mib \tau}_{{\bf R}'}) 
  (1 - 2 {\bf e}(\alpha _3) \cdot {\mib \tau}_{{\bf R}''}) ] ,  
  \end{array}
  \label{eq:eff3}
\end{equation}
where
$c={29 \over 384} J'^3/J^2$, $b={1 \over 384} J'^3/J^2$, 
${\bf e}(\alpha)$=$(\cos \alpha$,$\sin \alpha$,$0)$.  
Type-A tetrahedra constitute the fcc arrangement, and 
the summation is taken over all the tetrahedron 
triples of the second case explained above.  
Note that the total chirality is 
not conserved and also that 
the chirality pseudospins interact only in their transverse 
directions.  The angle $\alpha_j$ depends on the position 
of not only its corresponding tetrahedron but also 
the other two in each triple.  
Its value is shown in Fig.~3. 
The one-body terms in eq.~(\ref{eq:eff3}) are 
canceled out after taking the summation.  

\begin{figure}[tb]
$$\psboxto(8cm;0cm){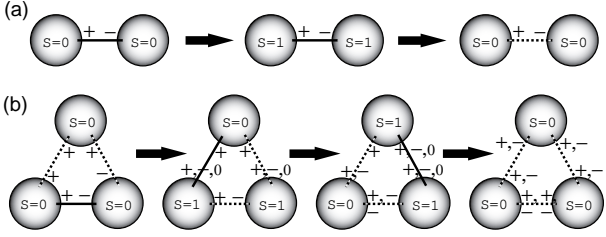}$$
\caption{Example of virtual processes in 
(a) the second and (b) the third order 
in $J'/J$.   Balls are type-A tetrahedra with 
chirality ($\tau$=$+,-,0$) indicated around 
each bond.  Perturbation in each step 
is indicated by a solid bond.} 
\label{fig4}
\end{figure}

We may understand the second and third-order effective models 
by examining their virtual processes.  It is essential to 
recall that the chirality of each spin-singlet eigenfunction 
on a tetrahedron is identical around any of its four 
sites.  On the other hand, a spin-triplet chiral eigenstate 
around one site is a mixture of 
all the three chiral eigenstates around another 
site.  It is also important that 
perturbation of a single bond conserves the 
chirality of each tetrahedron around the site to which 
the bond is connected, in both spin singlet and triplet multiplets.  
Therefore, as shown in Fig.~4(a), the chirality configuration 
on a tetrahedron pair always returns to the original one, 
after the second-order virtual processes, more generally 
in any orders, and the ground state degeneracy is not lifted.  
In the case of a coupled tetrahedron triple shown in Fig.~4(b), 
chirality is mixed up in the virtual processes including 
spin-triplet tetrahedra, since two bonds are connected to 
different sites on each tetrahedron.  Therefore, 
any of the eight chirality configurations 
can undergo a transition to all of them.  
This is why the effective 
model (\ref{eq:eff3}) has one-, two-, and three-body 
interactions in the transverse components in the chirality space. 
 
Now I discuss the ground state of the effective model (\ref{eq:eff3}) 
using the mean-field approach.  To this end, I first 
determine the mean-field ground state of its elementary 
building block shown in Fig.~3, the ``supertetrahedron'',  
each of its vertices ($j$=1$\sim$4) being a type-A tetrahedron.  
The mean-field Hamiltonian, $E$, is simply obtained by replacing 
chirality operators in eq.~(\ref{eq:eff3}) 
with classical unit vectors, 
${\mib \tau}$ $\rightarrow$
$(\cos \theta, \sin \theta, 0)$, 
and aside from a constant it is given as 
\begin{equation}
\begin{array}[t]{l}
 E(\theta_1,  \theta_2,  \theta_3,  \theta_4)= -b 
\left\{ 
(1 - 2 \cos \theta_2 ) 
[(1 - 2 \cos(\theta_3 + \phi)] 
\right. 
\\
\ \ 
\times 
[(1 - 2 \cos(\theta_4 - \phi)] 
+ (\theta_2,\theta_3,\theta_4 \rightarrow \theta_1,\theta_4,\theta_3) 
\\
\left.
\ \ 
+ (\theta_2,\theta_3,\theta_4 \rightarrow \theta_4,\theta_1,\theta_2) 
+ (\theta_2,\theta_3,\theta_4 \rightarrow \theta_3,\theta_2,\theta_1) 
\right\} .
\end{array}
\label{eq:MF}
\end{equation}
This has a symmetry associated with 
spatial rotation: 
$E(\theta_1, \theta_2, \theta_3, \theta_4)$= 
$E(\theta_1+\phi, \theta_3+\phi, \theta_4+\phi, \theta_2+\phi)$=
$E(\theta_4+\phi, \theta_2+\phi, \theta_1+\phi, \theta_3+\phi)$=
$E(\theta_2+\phi, \theta_4+\phi, \theta_3+\phi, \theta_1+\phi)$=
$E(\theta_3+\phi, \theta_1+\phi, \theta_2+\phi, \theta_4+\phi)$.  
I minimize this classical energy and find four types of 
ground-state configurations, 
\begin{equation}
(\theta_1,\theta_2,\theta_3,\theta_4) = 
\begin{array}[t]{l}
\displaystyle
(\theta_1,\pi,{\textstyle {1 \over 3}}\pi,{\textstyle {5 \over 3}}\pi ), \ 
(\pi,\theta_2,{\textstyle {5 \over 3}}\pi,{\textstyle {1 \over 3}}\pi), \ 
\\
\displaystyle
({\textstyle {1 \over 3}}\pi,{\textstyle {5 \over 3}}\pi,\theta_3,\pi), \ 
({\textstyle {5 \over 3}}\pi,{\textstyle {1 \over 3}}\pi,\pi,\theta_4). \ 
\end{array}
  \label{eq:conf}
\end{equation}
In each solution, one angle variable is still free 
and the classical energy is degenerate with respect to it.  

\begin{figure}[tb]
$$\psboxto(0cm;4cm){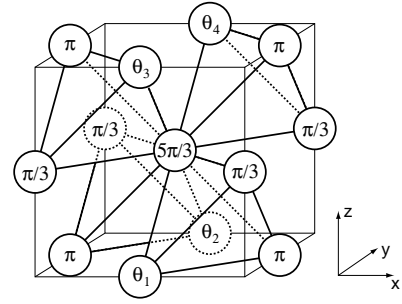}$$
\caption{Mean-field solution of the effective model 
in the chirality space.  Azimuthal angle $\theta$ of 
the ordered pseudospin is shown.} 
\label{fig5}
\end{figure}

I can patch up these four types of solutions 
for a supertetrahedron to construct a ground-state configuration 
on the entire fcc lattice, and the result is shown in Fig.~5.  
Three of the four sublattices show a long-range order.  
The other one remains disordered, and 
its residual entropy is ${1 \over 16} \log 2$ per original spin.  
I emphasize here that this solution is the global energy minimum 
within the mean-field approximation. 
This is because the solution minimizes the energy for 
each supertetrahedron in the whole fcc lattice 
without any extra energy cost.  
The degeneracy of this state is $4 \times 2^{N_{\rm A}/4}$, 
where the first factor is from the number of sublattices 
and the second one is from the disordered sublattice.  

The long-range order contributes 
to the chirality structure factor a set of Bragg 
peaks at the reciprocal lattice vectors of the simple cubic lattice, 
${\bf Q}=2\pi(n_x,n_y,n_z)$ ($n$'s are integer), in addition 
to a constant background due to the disordered sublattice.  
The amplitude of the Bragg peaks is given by 
$S_{\tau}({\bf Q})$=$N_A^{-2}$$\sum_{{\bf R},{\bf R}'}$
$e^{i{\bf Q}\cdot({\bf R}-{\bf R}')}$
$\langle {\mib \tau}_{\bf R}\cdot{\mib \tau}_{\bf R}' \rangle$
=${1 \over 8}$$(C_{yz}+C_{zx}+C_{xy})$, 
where $C_{yz}=\sin ^2 ({1 \over 2}(Q_y-Q_z))$=
${1 \over 2} (1-(-1)^{n_y+n_z})$ and so on, and therefore 
vanishes at ${\bf Q}$ where 
$n_x$, $n_y$, and $n_z$ have the same parity.  

The mean-field ground state has a finite 
expectation value of the pseudospin,  
$\langle {\mib \tau}_{\bf R} \rangle = (\cos \theta, \sin \theta, 0)$. 
However, its $z$-component describing spin chirality is zero, 
meaning that the time-reversal symmetry is not broken 
and dominant spin-spin correlations are collinear rather than chiral. 
The transverse components are finite instead, and 
I now discuss its physics with considering the wavefunction 
of four spins on an original tetrahedron. 
This for the given value of angle $\theta$ is given as 
the eigenfunction of 
the operator, $\langle {\mib \tau} \rangle \cdot {\mib \tau} 
= \tau^x  \cos \theta+ \tau^y \sin \theta$.  
All of the three cases, 
$\theta$=$\pi$,${1 \over 3} \pi$,${5 \over 3} \pi$, 
correspond to a valence bond state, i.e., a dimer pair, 
but with different pair combinations: 
\begin{equation}
  \theta = \pi : 
   [12] \otimes [34], \ \ 
  {\textstyle {1 \over 3}}\pi : 
  [13] \otimes [24], \ \ 
  {\textstyle {5 \over 3}}\pi : 
  [14] \otimes [23], \ \ 
  \label{eq:dimer}
\end{equation}
where $[ii']$ denotes  
the spin singlet wavefunction of two spins, $i$ and $i'$ (see Fig.~1).  
Therefore, the mean-field ground state has a long-range 
order of dimerization in three of the four tetrahedron sublattices and 
breaks the translation, mirror, and rotation symmetries in space. 
However, it does not break symmetry in spin space, 
since the order occurs within the spin-singlet multiplet.  
The possibility of dimerization was first discussed by Harris 
{\it et al.}~\cite{Harris}, 
and the present solution agrees with their proposal.  

Koga and Kawakami~\cite{Koga} also discussed dimerization on the 
pyrochlore lattice with including additional distortion 
which favors the ${\bf q}={\bf 0}$ order of either a tetramer 
singlet or one of the dimer pairs in eq.~(\ref{eq:dimer}), and 
found that these two states become degenerate 
as the distortion is switched off. 
Their solution is not continuously connected to the 
present solution, but this is because the distortion under consideration 
is limited to intra-tetrahedron modes with ${\bf q}={\bf 0}$.  
As the result of $S_{\tau}({\bf Q})$ shows, 
the present solution has a triple-${\bf q}$ structure 
with wavevectors ($2\pi$,0,0), (0,$2\pi$,0), and (0,0,$2\pi$).  
The dimerized state resonates with the tetramer state 
to realize the other two dimerized states in eq.~(\ref{eq:dimer}) 
with spatially modulated coefficients, 
and the three-body interactions in the effective model 
(\ref{eq:eff3}) stabilize the triple-${\bf q}$ structure 
with four sublattices.  

To compare with the experiments, one useful quantity 
is the spin structure factor, 
$S_{\sigma}({\bf q})$= 
$(4N_A)^{-1}$$\sum_{{\bf r},{\bf r}'}$
$e^{i{\bf q}\cdot({\bf r}-{\bf r}')}$
$\langle {\bf S}_{\bf r}\cdot{\bf S}_{\bf r}' \rangle$.
This is calculated 
for the mean-field ground state and the result is 
\begin{equation}
  S_{\sigma}({\bf q}) = {\textstyle {1 \over 4}} 
  [3-
  \begin{array}[t]{l}
  ( \cos {\textstyle {1 \over 4}} q_y \cos {\textstyle {1 \over 4}} q_z 
  + \cos {\textstyle {1 \over 4}} q_z \cos {\textstyle {1 \over 4}} q_x \\
  + \cos {\textstyle {1 \over 4}} q_x \cos {\textstyle {1 \over 4}} q_y )] . 
  \end{array}
  \label{eq:struct}
\end{equation}
Since this state is a spin singlet with no magnetic long range order, 
the spin structure factor has no Bragg peak and 
continuously changes with the wavevector.  
It has a maximum value at ${\bf q}$'s where one of 
$\cos {1 \over 4} q_\mu$ ($\mu$=$x$,$y$,$z$) is +1 and 
another one is $-1$, 
e.g., ${\bf Q}_1$=$(4\pi,4\pi,0)$ and 
${\bf Q}_2$=$(4\pi,2\pi,0)$.  Therefore, these ${\bf q}$'s form 
three orthogonal sets of parallel lines, 
reflecting strong geometrical frustration. 
This result (\ref{eq:struct}) is 
qualitatively similar to the result, 
Fig.~4, of Canals and Lacroix,~\cite{Canals} 
but the maximum position is shifted as for ${\bf Q}_1$ 
in their calculation.  
On the one hand, 
this may be because spins on different tetrahedra have 
no correlations within the present approach.  
On the other hand, Canals and Lacroix used 
the second-order perturbation in the inter-tetrahedron 
interaction $J'$.  As I pointed out above, 
up to the second order,   
the $J'$-term induces no spin-spin correlations 
between different tetrahedra for the singlet ground-state 
multiplet.  Therefore, the finite shift from ${\bf Q}_1$ 
in their result is due to the contribution 
of excited-state multiplets at finite temperatures, 
and it will vanish with the temperature approaching zero.  
We could improve the present result by also including perturbation 
for the wavefunction, but it is beyond the scope 
of this paper.  

In this paper, I have examined the ground state of 
the $S$=${1 \over 2}$ Heisenberg antiferromagnet on 
the pyrochlore lattice.  Assuming the ground state 
is a spin liquid, 
I obtained the effective model for a 
thermodynamic number of spin-singlet states 
below the spin gap and examined symmetry breaking, 
based on the third-order degenerate 
perturbation starting from decoupled tetrahedra.  
I determined its ground state using the mean-field 
approximation.  The result is a long-range order 
of spin-singlet dimers with a triple-${\bf q}$ structure, and 
three of the four constituent tetrahedron sublattices 
have a ordered dimer-pair pattern.  
Therefore, spin-triplet excitations have a finite 
energy gap, whereas spin-singlet excitations exist 
within the spin gap.  These singlet excitations are 
oscillations of the dimerization pattern in the ordered 
sublattices.  
One tetrahedron sublattice remains disordered, 
and it is randomly dimerized, tetramerized, even chiral, 
or any superposition of these.  
We expect that this sublattice will ultimately  
show a long-range order once quantum fluctuations 
are taken into account beyond the present approach.  
In a separate paper, I will discuss this problem 
both for spin and chirality degrees of freedom.  

I have investigated in the present study the lift of 
thermodynamic degeneracy 
driven by spin exchange processes, but there may exist 
other types of instability.  Recently, 
Yamashita and Ueda~\cite{Yamashita} found that a Jahn-Teller 
lattice distortion lifts this degeneracy 
for the $S$=1 antiferromagnet, and they proposed 
that this is a mechanism of a structural phase transition 
observed in ZnV$_2$O$_4$.~\cite{Ueda}  This compound also undergoes a 
magnetic phase transition at a lower temperature.  
To explain its magnetic order, theoretical approaches limited to 
the spin singlet multiplet are not sufficient and 
we must include finite-spin multiplets 
not only in the virtual processes but also in the real processes.  
This problem is left for a future project.  

I thank Frederic Mila for valuable discussions. 
This work was supported by a Grant-in-Aid from the
Ministry of Education, Science, Sports and Culture
and also by University of Tsukuba Research Projects.

\end{document}